# New diversity form of ice polymorphism: Discovery of second hydrogen ordered phase of ice VI

Ryo Yamane[1*], Kazuki Komatsu[2], Jun Gouchi[1], Yoshiya Uwatoko[1], Shinichi Machida[3], Takanori Hattori[4], Hayate Ito[2], Hiroyuki Kagi[2]

[1]The Institute for Solid State Physics, The University of Tokyo, 5-1-5 Kashiwanoha, Kashiwa, Chiba 277-8581, Japan,

[2]Geochemical Research Center, Graduate School of Science, The University of Tokyo, 7-3-1 Hongo, Bunkyo-ku, Tokyo 113-0033, Japan,

[3]Neutron Science and Technology Center, CROSS, 162-1 Shirakata, Tokai, Naka, Ibaraki 319-1106, Japan,

[4]J-PARC Center, Japan Atomic Energy Agency, 2-4 Shirakata, Tokai, Naka, Ibaraki 319-1195, Japan.

*Corresponding author
email: r.yamane@issp.u-tokyo.ac.jp

**More than 20 crystalline and amorphous phases have been reported for ice so far[1]. This extraordinary polymorphism of ice arises from the geometric flexibility of hydrogen bonds and hydrogen ordering[2], and makes ice a unique presence with its universality in the wide fields of material and earth and planetary science. A prominent unsolved question[3] concerning the diversity is whether a hydrogen-disordered phase of ice transforms into only one hydrogen-ordered phase, as inferred from the current phase diagram of ice, although its possible hydrogen configurations have close energies[4–7]. Recent experiments on a high-pressure hydrogen-disordered phase, ice VI, revealed an unknown hydrogen-ordered form (β-XV[8]) besides the known ordered phase, ice XV[9], which would be a counterexample of the question. However, due to lack of experimental evidence, it has not been clarified whether β-XV is a distinct crystalline phase[3,8,10–12]. Herein we report a second hydrogen-ordered phase for ice VI, ice XIX, unambiguously demonstrated by neutron diffraction measurements. The phase boundary between ice VI and ice XIX shows that ice VI contracts upon the hydrogen ordering, which thermodynamically stabilizes ice XIX in higher-pressure region than ice XV because of its smaller volume than ice XV[4,9,13]. The pressure-driven phase competition between hydrogen-ordered phases, also theoretically suggested in other ice polymorphs[6], can induce hydrogen ordering of ice in different manners. Thus, this study demonstrates a hitherto undiscovered polymorphism of ice.**

Comprehensive observation of hydrogen ordering in ice VI was conducted by dielectric experiments in the pressure range 0.88–2.2 GPa. Ice VI was initially obtained at room temperature and its dielectric properties were determined in both cooling and heating runs in the temperature range 100–150 K, using a newly developed pressure cell (see Methods). After the heating runs, the sample was subsequently heated to room temperature for annealing. Then, the sample was compressed again, and dielectric measurements were conducted at different pressures (Fig. 1).

Phase transitions from ice VI to its hydrogen-ordered phases were observed at around 120–130 K, along with sudden weakening of the dielectric response of ice VI with decreasing temperature (Fig. 2a and b). Hydrogen ordering of ice suppresses reorientation of water molecules which induces the dielectric response of ice[14,15]. We defined the disorder–order phase-transition temperatures from ice VI to its hydrogen-ordered phases as the starting temperature at which the slope ($dI/dT$; $I$: dielectric-loss peak intensity) changes (Fig. 2b). The slope of the obtained phase boundary, i.e. $dT/dP$, between ice VI and its hydrogen-ordered phases changes from negative to positive at around 1.6 GPa with increasing pressure (Fig. 1). Based on the Clausius–Clapeyron relationship, i.e. $dT/dP = \Delta V/\Delta S$, this sign change for $dT/dP$ strongly indicates that ice VI has two different hydrogen-ordered phases with opposite signs for $\Delta V$, because $\Delta S < 0$ generally holds for hydrogen ordering. Since the currently known hydrogen ordering from ice VI to ice XV shows a positive volume change (observed at the lower pressure, 0.4 GPa[4]), ice XV is in the lower pressure region and the hydrogen-ordered phase in the higher-pressure region is a new phase, ice XIX, which has a smaller volume than ice VI and also ice XV. The appearance of ice XIX is governed by the $PV$ term in the Gibbs energy expression, because the volume contraction thermodynamically stabilises ice XIX compared to ice XV. In this context, the phase boundary between ice XV and ice XIX should have a slope rather than lie horizontally as suggested previously[8,11], because ice XV has a larger volume than ice XIX (the supposed phase boundary in Fig. 1 is shown vertically to emphasise this point). It is noteworthy that the phase transition between ice VI and XIX showed hysteresis for the transition temperature (Extended data Fig. 1). This first-order phase transition is consistent with the sudden change in dielectric properties between ice VI and ice XIX (Fig. 2).

Neutron diffraction experiments were conducted at 1.6 and 2.2 GPa to confirm whether ice XIX is a hydrogen-ordered crystalline phase distinct from ice XV. Both cooling and heating runs were conducted at each pressure in the temperature range 80–150 K.

A transition from ice VI to ice XIX was also observed in the neutron diffraction experiments, as appearance of new peaks due to symmetry lowering (Fig. 3a). Some of the new peaks, e.g. those at 2.20 Å and 2.26 Å (indicated by blue triangles in Fig. 3a), cannot be assigned to the unit cell of ice XV; instead, they can be assigned to an expanded $\sqrt{2} \times \sqrt{2} \times 1$ cell with respect to the unit cell of ice VI (the unit cell of ice XV has a $1 \times 1 \times 1$ cell with respect to that of ice VI). This is unambiguous evidence that the hydrogen-ordered phase found in the higher-pressure region is a crystalline phase distinct from ice XV, and that ice VI has two different types of hydrogen ordering. The reflection conditions show that the unit cell of ice XIX has a primitive lattice. The reduced unit cell parameters of ice XIX, *a* and *c,* corresponding to the unit cell of ice VI, are expanded and contracted, respectively, upon hydrogen ordering (Fig. 3b); this tendency was also observed at 2.2 GPa. A comparison of the temperature dependences of *c*/*a* at 1.6 and 2.2 GPa (Fig. 3c) showed that the phase-transition temperature at 2.2 GPa was at about 7 K higher than that at 1.6 GPa. This result is consistent with the phase boundary between ice VI and XIX obtained by the dielectric experiments. On the other hand, no significant volume change was observed in our neutron diffraction experiments, in contrast to the expected negative volume change ($\Delta V < 0$) upon hydrogen ordering, probably due to the small volume contraction.

For the structure analysis of ice XIX, we considered candidates of its space group based on the group−subgroup relationship between ice VI and XIX, in addition to the experimentally confirmed reflection conditions. There are 36 subgroups for the space group of ice VI, *P*4$_2$/*nmc*, considering the primitive unit cell of ice XIX. Among them, thirteen space groups, having $h0l: h + l = 2n$ and $0kl: k + l = 2n$ reflection conditions, can be excluded from the observed reflection conditions. We conducted Rietveld analyses using structural models with 18 space groups of the remaining candidates, except for the lower-symmetry space groups: $Pc$, $P2_1$, $P2$, $P\bar{1}$ and $P1$—this cut-off is based on indices of the subgroups of $P4_2/nmc$ (see details in Supplementary). Notably, we do not rule out the possibility that the actual crystal structure of ice XIX having one of these space groups, although sufficient refinement agreements were obtained for the 18 candidates from our neutron diffraction data. A structural model of each candidate was constructed using a partially ordered model adopted in an earlier study[4]. $P\bar{4}$ or $Pcc2$ structural models are the most plausible for the space group of ice XIX, based on the structure refinements. Considering the suggested space group of ice XV, $P\bar{1}$[9] or $Pmmn$[4], centrosymmetry of hydrogen configurations is the most significance difference in hydrogen configuration between ice XIX and ice XV. In particular, $Pcc2$ suggests a pyroelectric structure as well as ice XI and its polar direction is along the $c$ axis. Although further investigations, such as a single-crystal neutron diffraction experiment, are necessary to precisely determine the hydrogen configurations, centrosymmetry will be an intriguing point in structural studies of ice XV and XIX.

Past arguments for the second hydrogen-ordered phase of ice VI should be mentioned here[3,8,10–12]. The existence of such a phase (β-XV[8]) in decompressed samples from above 1.45 GPa was first suggested by Gasser et al. using various measurements at ambient pressure[8]. For example, differential scanning calorimetry (DSC) gave an unassignable endotherm peak from the known phase transition between ice VI and XV. Rosu-Finsen et al.[10] reported further detailed DSC experiments conducted under ambient pressure at different cooling/heating rates and using different quenching/annealing procedures. Their DSC results clarified that the peak also appeared for quenched samples decompressed from 1.0 GPa, where ice XIX did not appear[8,10,11], and its appearance/disappearance depends on the heating rate used in the DSC experiment. Based on these results, they questioned the supposed existence of a second hydrogen-ordered phase[8], and suggested another scenario by introducing the idea of the deep-glassy state of ice VI to interpret the observed DSC profiles. Based on the in-situ observations presented herein, we firmly confirmed the presence of a second hydrogen-ordered phase of ice VI under high pressure, and named it ice XIX, as a crystalline phase distinct from ice XV. In our view for the past arguments, although the decompressed samples in the higher-pressure region should undergo hydrogen ordering from ice VI to ice XIX before decompression, it is necessary to investigate whether the decompressed samples retain the crystal structure of ice XIX as well as its crystallinity. This is because the idea of the deep-glassy state seems reasonable based on the DSC profiles of their decompressed samples, and on comparing our *in-situ* dielectric loss data to that measured under ambient pressure[8], revival of reorientation dynamics, which should be immobilised upon hydrogen ordering[15], was evidenced by a reappearance of dielectric loss of the decompressed samples under ambient pressure[8] (see also Extended Data Fig. 1b). This reactivated reorientation might partially break the long-range hydrogen order of ice XIX to obtain more stable configurations under lower pressure. Further investigation is necessary to ensure consistency among all the observed data, based on both the newly found ice XIX and the concept of the deep-glassy state.

This study first demonstrates the existence of multiple hydrogen-ordered phases for a hydrogen-disordered phase, and clarifies the effectiveness of applying pressure to induce phase competition among the hydrogen-ordered phases. Based on previous theoretical studies[4–7,16] and the currently known phase diagram of ice, the low-temperature region of the phase diagram (below approx. 150 K) is a frontier region for exploring undiscovered ways of hydrogen-ordering in ice, which would greatly change the phase diagram of ice. It is additionally noteworthy that the unit cell size of ice XIX allows many possible hydrogen-ordered configurations (1964 symmetry-independent configurations), such that an exhaustive theoretical analysis for the all configurations is difficult. However, such a wide variety of hydrogen-ordered configurations and their stability evaluations might be a good benchmark for modern theoretical trials toward modelling biochemical and environmental processes with large water molecules, such as using topological graph invariant theory[17], combining oriented graph theory and density functional calculations, which can evaluate the energy stability of a large number of water-molecule arrangements. To the best of our knowledge, this is also the first report of a hydrogen-bonded material for which different hydrogen-ordered configurations are realised depending on the pressure, although electric field is a known effective parameter to control ferro- and antiferroelectric structures of organic hydrogen-bonded ferroelectrics[18,19]. This newly discovered coupling between hydrogen bond and pressure will extensively develop a new research field focusing on the pressure-controllability of hydrogen-ordered configurations, which potentially include significant physical properties, e.g. piezo- and (anti)ferro-electric, using established techniques of neutron diffraction experiments.

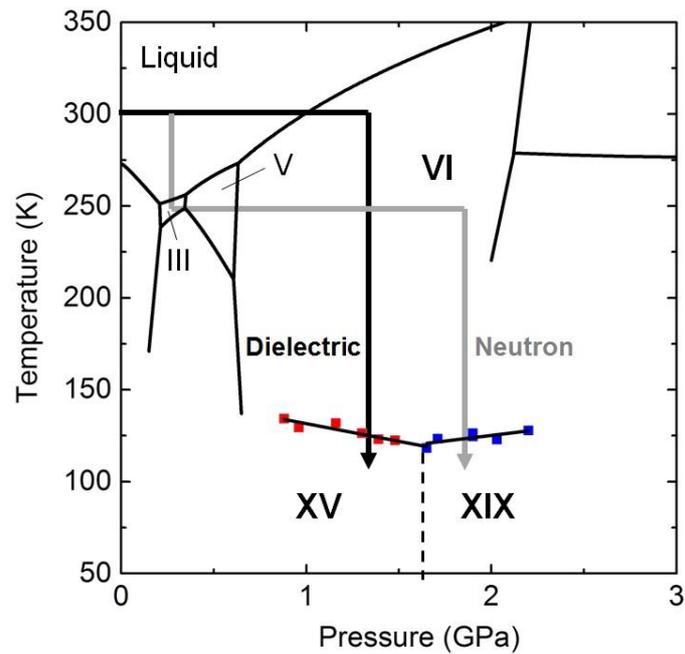

**Figure 1| Representative experimental paths of dielectric and neutron diffraction experiments described in the phase diagram of ice obtained herein.** Dielectric experiments of ice VI and its hydrogen-ordered phases were conducted at 0.88–2.2 GPa. HCl (99.9%, Wako) was introduced as a dopant (concentration: $10^{-2}$ M) to accelerate the hydrogen ordering of ice VI[20]. The measured temperature was in the range 100–150 K and changed at a rate of 2 K/h. Neutron diffraction experiments of DCl-doped $D_2O$ (concentration: $10^{-2}$ M) were conducted using a more complicated path to ensure that the sample was a fine powder through solid–solid phase transitions, i.e., ice III→ice V→ice VI. Sample diffraction was collected at 1.6 and 2.2 GPa, and the temperature range was 80–150 K. Temperature was changed at a rate of 6 K/h. Diffraction patterns were collected using new samples in each run at different pressures to confirm reproducibility. Phase boundaries among ice VI, ice XV, and ice XIX are described by black solid lines, based on dielectric experiments (red and blue squares correspond to phase transition temperatures from ice VI to ice XV and XIX, respectively). The dotted line shows the provisional phase boundary between ice XV and ice XIX (see main text).

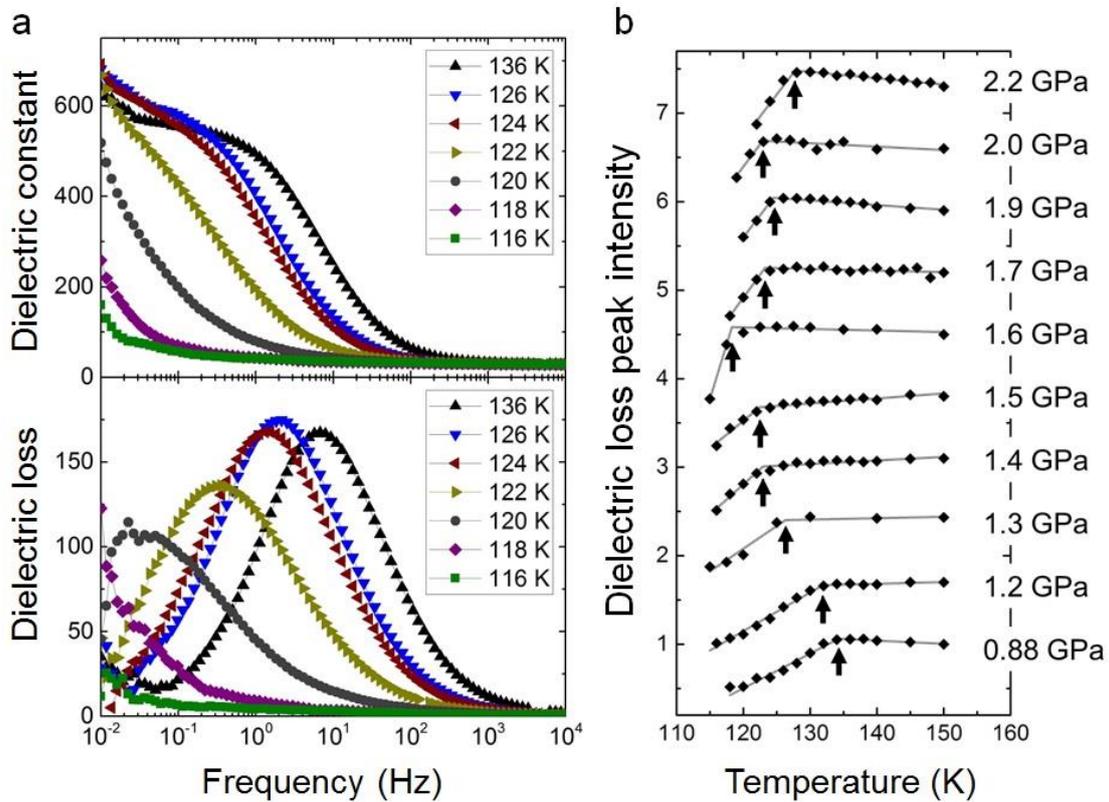

**Figure 2| Temperature dependence of dielectric properties of HCl-doped ice VI and its hydrogen-ordered phases. a**, Dielectric constant and dielectric loss of HCl-doped ice VI and its hydrogen-ordered phase (ice XIX) obtained at 1.9 GPa upon cooling. The measured frequency was from 3 mHz to 2 MHz. **b**, Temperature dependence of dielectric loss peak intensity of HCl-doped ice VI and its hydrogen-ordered phases obtained in the pressure range 0.88–2.2 GPa upon cooling (black diamonds). Each peak intensity of dielectric loss was estimated using a model fitting the corresponding dielectric loss spectrum based on the Debye dielectric-relaxation equation (polydispersion type). Under each pressure, peak intensities were normalised by that obtained at the highest temperature. Each plot was shifted by 0.7 with increasing pressure for clarity. The grey lines were separately fitted for the data of loss peak intensity originating from ice VI and its hydrogen-ordered phases at each pressure. The phase-transition temperature was defined by an intersection of the two grey lines. The black arrows indicate the phase transition temperature at each pressure.

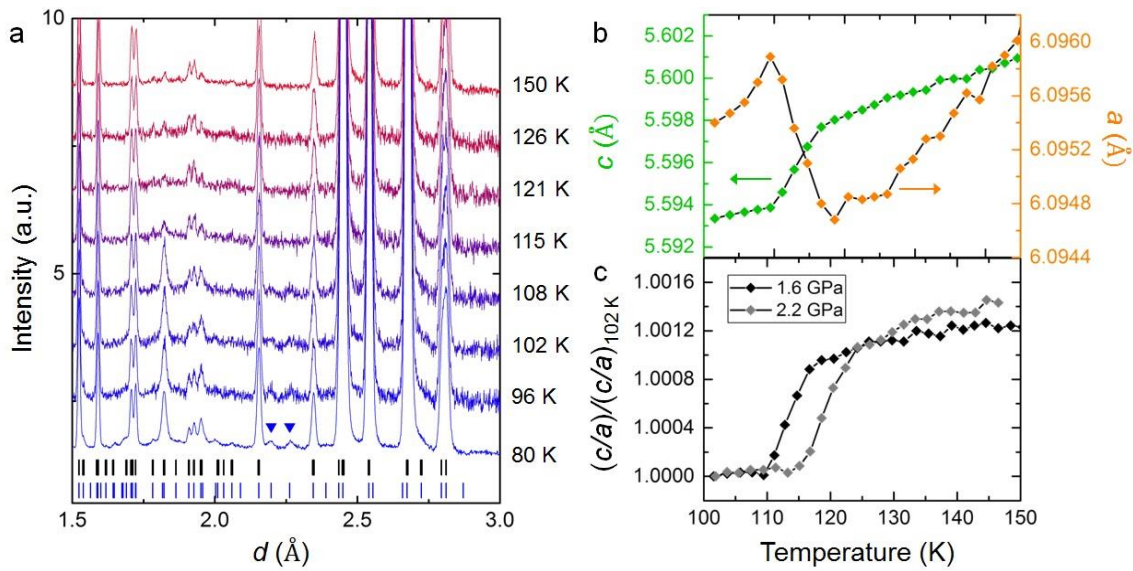

**Figure 3| Temperature dependence of neutron diffraction patterns and lattice parameters of DCl-doped $D_2O$ ice VI and ice XIX. a**, Neutron diffraction patterns of ice VI and XIX obtained at 1.6 GPa in the cooling run. Only an expanded area showing new peaks of ice XIX is displayed. The blue and black ticks represent all the peak positions expected from the unit cells of ice XIX and ice XV, respectively. Blue triangles indicate new peaks at 2.20 Å and 2.26 Å, which do not appear from the unit cell of ice XV. **b**. Temperature dependence of lattice parameters, $a$ and $c$, of ice VI or ice XIX obtained at 1.6 GPa. The values were calculated based on the ice VI structure model even for ice XIX because of a common oxygen framework between ice VI and XIX. **c**. Temperature dependence of $c/a$ at two different pressures, 1.6 and 2.2 GPa, indicated by black and grey. Phase transition from ice VI to ice XIX started at around 117 K and 124 K in the respective cooling runs. The $c/a$ values are normalised by that at 102 K. Diffraction patterns were collected using new samples in each run under different pressures to confirm reproducibility.

# References

1. Millot, M. *et al.* Nanosecond X-ray diffraction of shock-compressed superionic water ice. *Nature* **569,** 251–255 (2019).
2. Salzmann, C. G., Radaelli, P. G., Slater, B. & Finney, J. L. The polymorphism of ice: Five unresolved questions. *Phys. Chem. Chem. Phys.* **13,** 18468–18480 (2011).
3. Salzmann, C. G. Advances in the experimental exploration of water's phase diagram. *J. Chem. Phys.* **150,** 60901 (2019).
4. Komatsu, K. *et al.* Partially ordered state of ice XV. *Sci. Rep.* **6,** 1–12 (2016).
5. Knight, C. & Singer, S. J. Hydrogen bond ordering in ice v and the transition to ice XIII. *J. Chem. Phys.* **129,** 164513 (2008).
6. Tribello, G. A., Slater, B. & Salzmann, C. G. A blind structure prediction of ice XIV. *J. Am. Chem. Soc.* **128,** 12594–12595 (2006).
7. Del Ben, M., Vandevondele, J. & Slater, B. Periodic MP2, RPA, and boundary condition assessment of hydrogen ordering in ice XV. *J. Phys. Chem. Lett.* **5,** 4122–4128 (2014).
8. Gasser, T. M. *et al.* Experiments indicating a second hydrogen ordered phase of ice VI. *Chem. Sci.* **9,** 4224–4234 (2018).
9. Salzmann, C. G., Radaelli, P. G., Mayer, E. & Finney, J. L. Ice XV: A new thermodynamically stable phase of ice. *Phys. Rev. Lett.* **103,** 105701 (2009).
10. Rosu-Finsen, A. & Salzmann, C. G. Origin of the low-temperature endotherm of acid-doped ice VI: New hydrogen-ordered phase of ice or deep glassy states? *Chem. Sci.* **10,** 515–523 (2019).
11. Thoeny, A. V., Gasser, T. M. & Loerting, T. Distinguishing ice β-XV from deep glassy ice VI: Raman spectroscopy. *Phys. Chem. Chem. Phys.* **21,** 15452–15462 (2019).
12. Rosu-Finsen, A., Amon, A., Armstrong, J., Fernandez-Alonso, F. & Salzmann, C. G. Deep-glassy ice VI revealed with a combination of neutron spectroscopy and diffraction. *J. Phys. Chem. Lett.* **11,** 1106–1111 (2020).
13. Salzmann, C. G. *et al.* Detailed crystallographic analysis of the ice VI to ice XV hydrogen ordering phase transition. *J. Chem. Phys.* **145,** 204501 (2016).
14. Kawada, S. Dielectric dispersion and phase transition of KOH doped Ice. *J. Phys. Soc. Japan* **32,** 1442 (1972).
15. Whalley, E., Davidson, D. W. & Heath, J. B. R. Dielectric properties of ice VII. Ice VIII: A new phase of ice. *J. Chem. Phys.* **45,** 3976–3982 (1966).
16. Singer, S. J. *et al.* Hydrogen-bond topology and the ice VII/VIII and ice Ih/XI proton-ordering phase transitions. *Phys. Rev. Lett.* **94,** 1–4 (2005).
17. Mcdonald, S., Ojamae, L. & Singer, S. J. Graph theoretical generation and analysis of hydrogen-bonded structures with applications to the neutral and protonated Water cube and dodecahedral clusters. *J. Phys. Chem. A* **5639,** 2824–2832 (1998).
18. Horiuchi, S. & Ishibashi, S. Hydrogen-bonded small-molecular crystals yielding strong

ferroelectric and antiferroelectric polarizations. *J. Phys. Soc. Japan* **89,** 51009 (2020).

19. Horiuchi, S., Kumai, R. & Ishibashi, S. Strong polarization switching with low-energy loss in hydrogen-bonded organic antiferroelectrics. *Chem. Sci.* **9,** 425–432 (2018).
20. Salzmann, C. G., Radaelli, P. G., Hallbrucker, A., Mayer, E. & Finney, J. L. The preparation and structures of hydrogen ordered phase of ice. *Phys. Chem. Ice* **311,** 521–528 (2006).
21. Hattori, T. *et al.* Design and performance of high-pressure PLANET beamline at pulsed neutron source at J-PARC. *Nucl. Inst. Methods Phys. Res. A* **780,** 55–67 (2015).
22. Komatsu, K. *et al.* Development of a new $P-T$ controlling system for neutron-scattering experiments. *High Press. Res.* **33,** 208–213 (2013).
23. Strassle, T., Klotz, S., Kunc, K., Pomjakushin, V. & White, J. S. Equation of state of lead from high-pressure neutron diffraction up to 8.9 GPa and its implication for the NaCl pressure scale. *Phys. Rev. B* **90,** 14101 (2014).
24. Aroyo, M. I., Kirov, A., Capillas, C., Perez-Mato, J. M. & Wondratschek, H. Bilbao Crystallographic Server. II. Representations of crystallographic point groups and space groups. *Acta Crystallogr. Sect. A Found. Crystallogr.* **62,** 115–128 (2006).
25. Mois Ilia Aroyo, Juan Manuel Perez-Mato, Cesar Capillas, Eli Kroumova, Svetoslav Ivantchev, Gotzon Madariaga, Asen Kirov, H. W. Bilbao Crystallographic Server: I. Databases and crystallographic computing programs. *Zeitschrift fur Krist.* **221,** 15–27 (2006).
26. Aroyo, M. I. *et al.* Crystallography online: Bilbao crystallographic server. *Bulg. Chem. Commun.* **43,** 183–197 (2011).
27. Toby, B. H. EXPGUI, a graphical user interface for GSAS. *J. Appl. Crystallogr.* **34,** 210–213 (2001).
28. Larson, A. C. & Von Dreele, R. B. *GSAS: General Structure Analysis System.* (2004).

#**Acknowledgement**

We are grateful to the technical staff of the University of Tokyo (Graduate School of Science), Mr. S. Otsuka and Mr. T. Shimozawa, for their support in the experiments. Neutron diffraction experiments were performed using the J-PARC user program (proposal number 2019A0310). This research was supported by JSPS KAKENHI (Grant numbers: 19H00648, 18J13298, 18H05224, 18H01936, 15H05829).

#**Author contribution**

R.Y. conceived and designed the experiments. R.Y., J.G., and Y.U. developed the high-pressure cell for dielectric measurements. R.Y., J.G., and H.I conducted the dielectric experiments. R.Y., K.K., S.M., and T.H. conducted the neutron diffraction experiments. R.Y. and K.K. analysed the neutron diffraction data. R.Y. wrote the manuscript with contributions from K.K., T.H., and H.K. All the authors have discussed the data interpretation.

## Methods

### Dielectric measurements

We conducted *in-situ* dielectric measurements under high pressure using a newly developed cell assembly. One of the most notable features of our development is that along with measuring the dielectric properties of the sample, the sample pressure can be simultaneously estimated using the ruby fluorescence method. This feature allows us to closely investigate the phase structure of ice in terms of its hydrogen ordering. The cell assembly is based on a piston-cylinder-type high-pressure apparatus (see details of the cell assembly in Supplementary Information). In the dielectric experiments involving ice VI, HCl (99.9%, Wako) was introduced as a dopant (concentration: $10^{-2}$ M) to accelerate the hydrogen ordering of ice VI[20], and dielectric experiments were conducted on DCl-doped $D_2O$ ice VI (DCl concentration: $10^{-2}$ M) following the same experiment procedure as that for HCl-doped ice VI (Fig. 1). Pressure dependence of the phase transitions was similar to that of the HCl-doped $H_2O$ sample, although the stable pressure region of the new hydrogen-ordered phase seemed to expand slightly to a lower-pressure region (between 1.3 and 1.5 GPa, see Extended Data Fig. 2).

### Neutron diffraction measurements

The neutron diffraction measurements were conducted at PLANET beamline 11 at the Materials and Life Science Experimental Facility of J-PARC, Ibaraki, Japan[21]. DCl-doped $D_2O$ sample was used as a starting material (DCl concentration: $10^{-2}$ M), and ice VI was prepared through solid–solid phase transitions, ice III→V→VI, to obtain a fine powder sample (Fig. 1). Pressure and temperature were controlled by using a Mito-system[22], and the pressure was estimated from the lattice parameter of Pb, which was added to the sample as a pressure marker[23].

### Structure analysis of ice XIX

Initial candidates of ice XIX were determined based on the group–subgroup relationship between ice

VI and XIX, using the SUBGROUPS program opened on the Bilbao crystallographic server[24–26]. The structural models for the 18 space group candidates were constructed based on the partially hydrogen-ordered model adopted in a previous study[4]. Hydrogen occupancies and atomic coordinates were the fitting parameters obeying the ice rule in this model (details of the 18 structural models are given in Supplementary Information). Structure refinements were conducted for the neutron diffraction patterns corrected at 1.6 GPa and 80 K using all the structure models employing the Rietveld method (program: GSAS with EXPGUI[27,28]). We used the initial structural parameters with the ice VI structural model, based on the neutron diffraction pattern obtained at 1.6 GPa and 80 K. It was noted that site occupancies of hydrogen atoms were initially considered fitting parameters in the refinements with fixed atomic positions of hydrogen. $\chi^2$ values of the structure models are plotted in Extended Data Fig. 3, where numerical values are shown only for five candidates with $\chi^2 <$ 9. For each structural model, Rietveld refinements were performed several times to confirm their reproducibility. The space group $P\bar{4}$ was deemed the most plausible candidate in this step. Extended Data Fig. 4 shows fitted lines using the five possible structural models (red-coloured) for the neutron diffraction pattern obtained at 1.6 GPa and 80 K (black lines). It should be mentioned that an observed Bragg peak, marked by a black tick at 1.82 Å, shows broadening compared to the simulated ones. This peak is derived from the new hydrogen-ordered phase. Generally, peak broadening arises from two factors, insufficient crystallite size and/or microstrain in the crystal, consistent with the partially hydrogen-ordered state of ice XIX. Structure refinements, including atomic positions of hydrogen, were conducted for the five candidates (Extended Data Fig. 5); the fitted results obtained employing only hydrogen occupancies as fitting parameters (Extended Data Fig. 4) were subsequently used for these refinements. $\chi^2$ values of $P\bar{4}$ and $Pcc2$ were comparable considering the dispersion of their refinement results, and those structural models were considered the most plausible for ice XIX. Atomic fractional coordinates for ice XIX using the two models are shown in Extended Data Tables 1 and 2, where lattice parameters have also been refined.

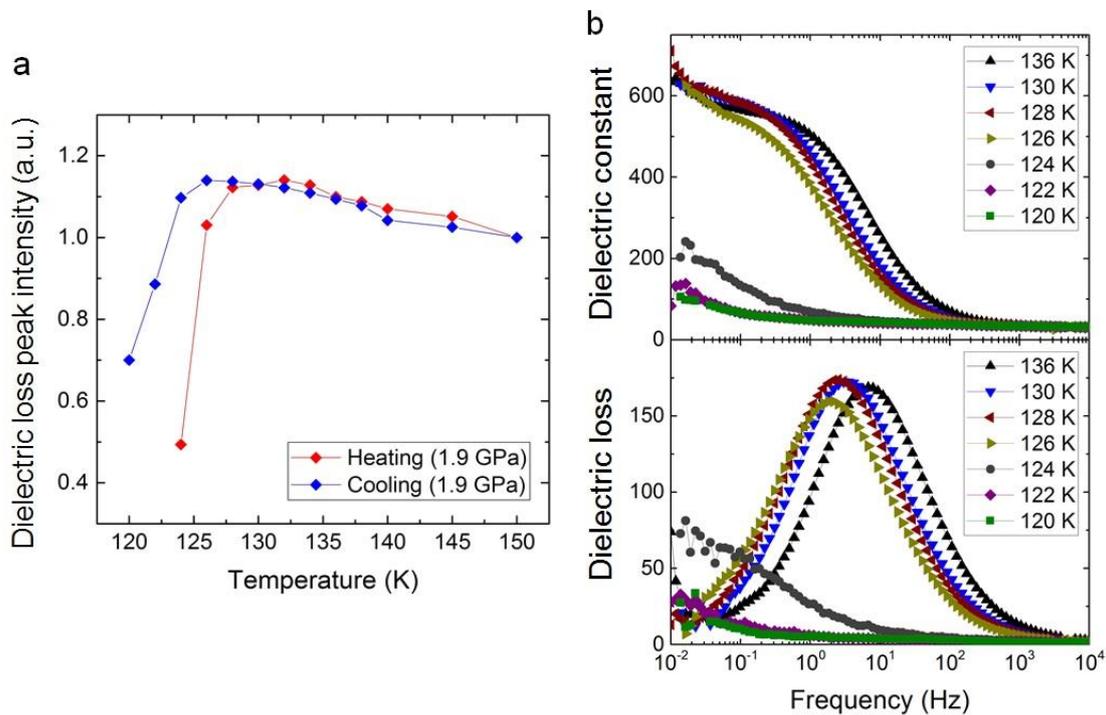

**Extended data Figure 1| Comparison of the phase transition in cooling and heating runs at 1.9 GPa. a,** Each peak intensity of dielectric loss is normalised by that obtained at the highest temperature in each run. At temperatures lower than the lowest temperature points shown in the figure, the dielectric response of ice XIX almost disappeared in the measured frequency region; thus, the peak intensity in that region is not shown here. Each of the cooling and heating results corresponds to the data shown in Fig. 2a (main text) and this Extended Data Fig. 1b, respectively. **b,** Dielectric constant and dielectric loss of HCl-doped ice VI and its hydrogen-ordered phase (ice XIX) obtained at 1.9 GPa upon heating. The measured frequency ranged from 3 mHz to 2 MHz.

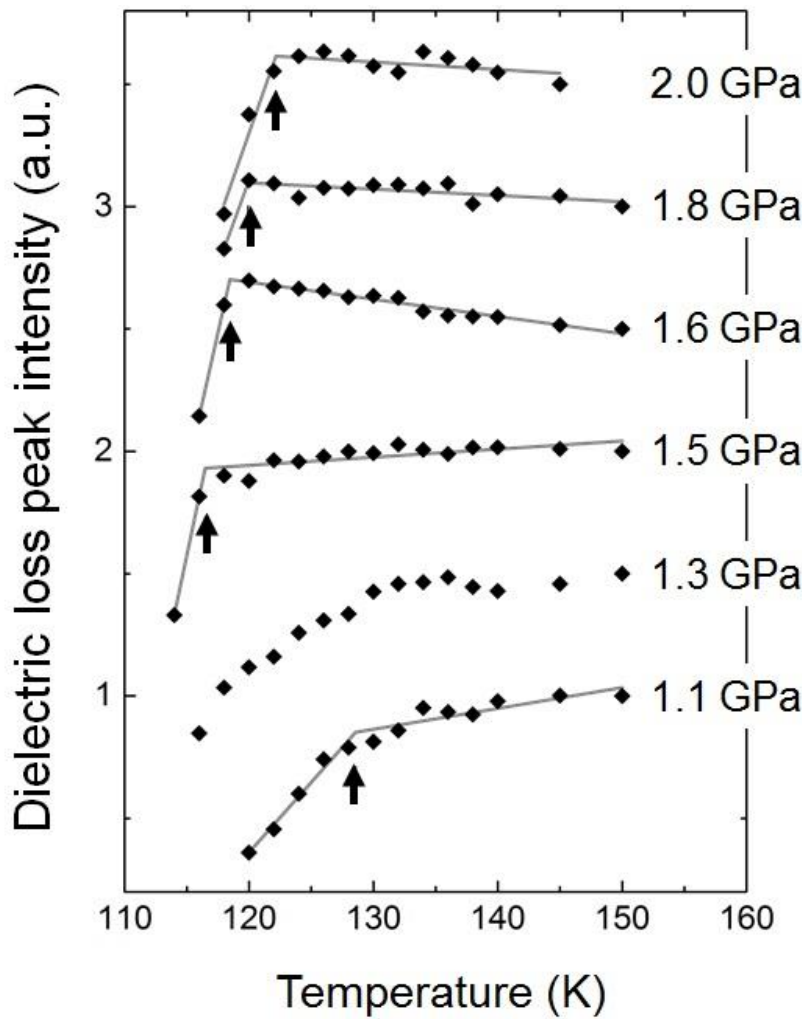

**Extended data Figure 2| Temperature dependence of dielectric loss peak intensity of DCl-doped $D_2O$ ice VI and its hydrogen ordered phases obtained from 1.1 to 2.0 GPa.** Peak intensities are normalised by that obtained at the highest temperature in each pressure run. Each plot was shifted by 0.5 with increasing measured pressure for clarity. The black diamonds represent obtained data. The grey lines were separately fitted to the data of loss peak intensity originating from ice VI and its hydrogen-ordered phases at each pressure (Fitted lines for 1.3 GPa are not shown due to its ambiguous change in the slope). The black arrow indicates the phase transition temperature at each pressure.

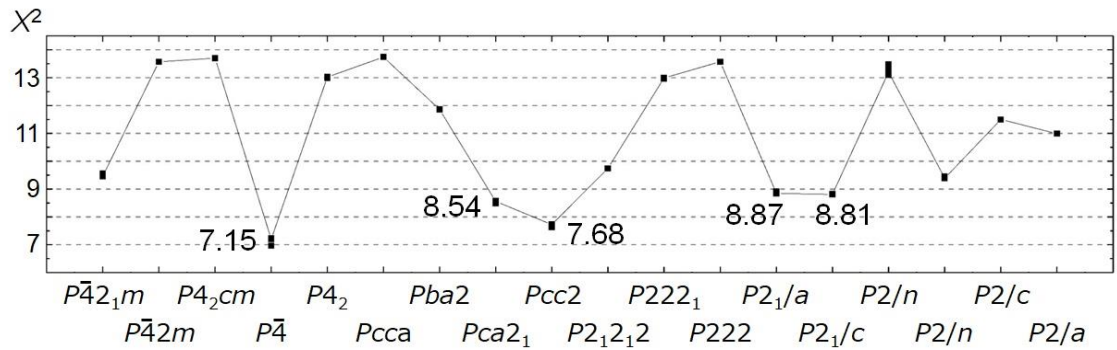

**Extended data Figure 3| Comparison of $\chi^2$ values obtained by Rietveld analysis for the neutron diffraction pattern obtained at 1.6 GPa and 80 K using the 18 structure models.** Structure refinements were conducted several times for each model to confirm their reproducibility; these results are plotted in this figure. The numerical values are shown only for five candidates with $\chi^2 < 9$, which are averaged values over several refinement results. Site occupancies of hydrogen atoms were initially refined by fixing the atomic positions of hydrogen.

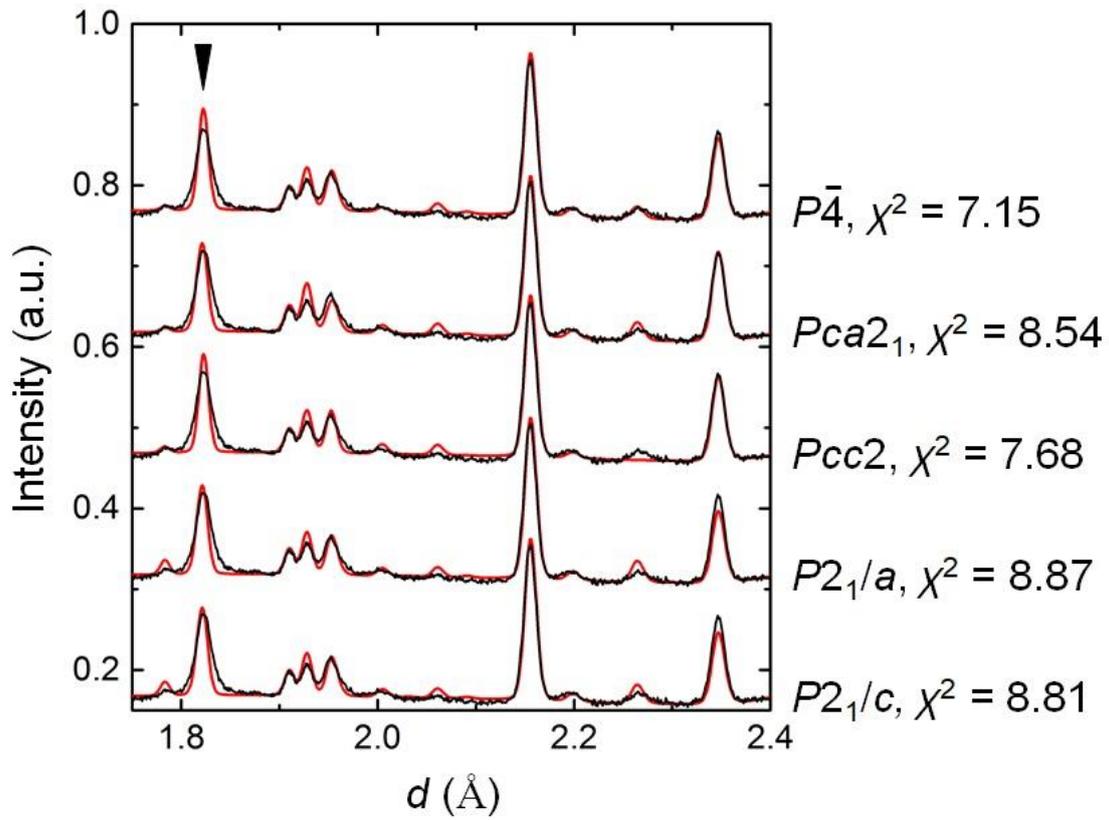

**Extended data Figure 4| Comparison of structure refinements of the five possible candidates.** Black lines show the neutron diffraction patterns observed at 1.6 GPa and 80 K. The results of the simulation for each model are shown by red lines. Space groups and $\chi^2$ values of the models are listed on the right-hand side of this figure. The black triangle indicates a Bragg peak that shows peak broadening compared to the simulated results.

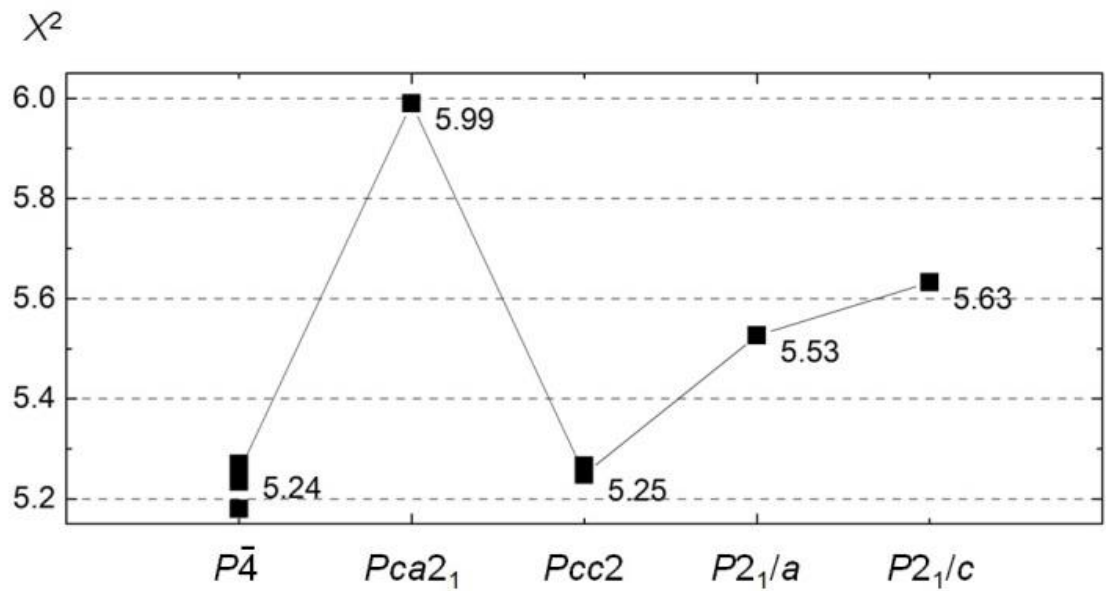

**Extended data Figure 5| Comparison of $\chi^2$ values obtained by Rietveld analysis for neutron diffraction pattern at 1.6 GPa and 80 K using the five plausible structure models.** In the refinements, atomic positions of hydrogen as well as the site occupancies of hydrogen atoms are refined. The $\chi^2$ values shown for each structure model is the averaged value obtained over repeated refinements.

Extended Data Table 1| Atomic fractional coordinates for ice XIX using $P\bar{4}$ (tetragonal) structure model (lattice parameters: $a = b = 8.61941(7)$ and $c = 5.59297(8)$)

| Atom | x | y | z | Occupancy |
| --- | --- | --- | --- | --- |
| O1a | 0.0 | 0.5 | 0.007(3) | 1.0 |
| O1b | 0.0 | 0.0 | 0.5 | 1.0 |
| O1c | 0.5 | 0.5 | 0.5 | 1.0 |
| O2a | 0.1436(15) | 0.6429(18) | 0.368(2) | 1.0 |
| O2b | 0.3663(19) | 0.8575(19) | 0.376(2) | 1.0 |
| O2c | 0.8660(19) | 0.8480(14) | 0.116(3) | 1.0 |
| O2d | 0.635(2) | 0.6411(18) | 0.111(3) | 1.0 |
| D1a | 0.116(5) | 0.593(5) | 0.269(6) | 0.324(14) |
| D1b | 0.895(3) | 0.904(4) | 0.260(6) | 0.5 |
| D1c | 0.394(3) | 0.903(3) | 0.226(4) | 0.676(14) |
| D1d | 0.602(4) | 0.611(3) | 0.282(6) | 0.5 |
| D2a | 0.061(3) | 0.565(3) | 0.103(3) | 0.676(14) |
| D2b | 0.939(4) | 0.939(4) | 0.415(6) | 0.5 |
| D2c | 0.442(5) | 0.959(5) | 0.138(6) | 0.324(14) |
| D2d | 0.547(4) | 0.565(4) | 0.393(7) | 0.5 |
| D3a | 0.221(4) | 0.719(4) | 0.376(4) | 0.55(3) |
| D3b | 0.782(4) | 0.784(5) | 0.128(8) | 0.5 |
| D3c | 0.283(4) | 0.780(5) | 0.371(5) | 0.45(3) |
| D3d | 0.723(4) | 0.711(4) | 0.122(8) | 0.5 |
| D4a | 0.833(3) | 0.952(3) | 0.039(4) | 0.69(2) |
| D4b | 0.177(5) | 0.549(5) | 0.450(6) | 0.40(2) |
| D4c | 0.666(2) | 0.551(3) | 0.034(4) | 0.78(2) |
| D4d | 0.308(5) | 0.934(4) | 0.414(6) | 0.27(7) |
| D4e | 0.950(3) | 0.333(3) | 0.466(4) | 0.73(7) |
| D4f | 0.544(8) | 0.672(8) | 0.031(11) | 0.22(2) |
| D4g | 0.550(3) | 0.168(4) | 0.469(5) | 0.60(2) |
| D4h | 0.954(3) | 0.877(3) | 0.010(5) | 0.32(2) |

Extended Data Table 2| Atomic fractional coordinates for ice XIX using $Pcc2$ (orthorhombic) structure model (lattice parameters: $a = 8.6111(4)$, $b = 8.6280(4)$, and $c = 5.59302(8)$)

| Atom | x | y | z | Occupancy |
| --- | --- | --- | --- | --- |
| O1 | 0.2536(15) | 0.752(2) | 0.760(3) | 1.0 |
| O2a | 0.3846(17) | 0.8931(17) | 0.133(3) | 1.0 |
| O2b | 0.1037(15) | 0.1150(16) | 0.869(5) | 1.0 |
| O2c | 0.1108(18) | 0.3905(17) | 0.623(3) | 1.0 |
| O2d | 0.6070(16) | 0.6066(17) | 0.8698 | 1.0 |
| D1a | 0.355(3) | 0.865(3) | 0.989(5) | 0.62(2) |
| D1b | 0.145(5) | 0.159(4) | 0.030(6) | 0.42(3) |
| D1c | 0.139(4) | 0.334(3) | 0.501(7) | 0.44(3) |
| D1d | 0.640(4) | 0.651(3) | 0.022(4) | 0.52(3) |
| D2a | 0.310(4) | 0.800(4) | 0.867(8) | 0.38(2) |
| D2b | 0.189(3) | 0.191(3) | 0.145(5) | 0.58(3) |
| D2c | 0.204(3) | 0.316(3) | 0.372(5) | 0.56(3) |
| D2d | 0.682(3) | 0.687(3) | 0.178(5) | 0.48(2) |
| D3a | 0.475(3) | 0.967(3) | 0.129(6) | 0.5 |
| D3b | 0.030(3) | 0.028(3) | 0.871(4) | 0.5 |
| D3c | 0.037(3) | 0.469(3) | 0.616(4) | 0.5 |
| D3d | 0.540(3) | 0.528(3) | 0.877(5) | 0.5 |
| D4a | 0.065(4) | 0.192(5) | 0.831(6) | 0.29(2) |
| D4b | 0.413(3) | 0.790(3) | 0.221(6) | 0.67(2) |
| D4c | 0.809(3) | 0.076(2) | 0.285(5) | 0.79(3) |
| D4d | 0.199(5) | 0.431(4) | 0.713(8) | 0.35(3) |
| D4e | 0.586(5) | 0.708(6) | 0.789(9) | 0.33(2) |
| D4f | 0.917(2) | 0.303(2) | 0.221(5) | 0.71(2) |
| D4g | 0.300(3) | 0.589(2) | 0.291(5) | 0.65(3) |
| D4h | 0.709(4) | 0.8577 | 0.726(8) | 0.21(3) |

**Supplementary information**

**Supplementary Methods: Dielectric measurements**

Dielectric experiments were conducted using a newly developed cell assembly for *in-situ* dielectric measurements under high pressure (Supplementary Figure 1). The cell assembly is based on piston-cylinder type high-pressure apparatus. The left side of Supplementary Figure 1 shows an overall of the piston-cylinder cell. The Developed cell assembly is shown on the right side of Supplementary Figure S1. A sample is loaded along to the vertical direction, and electric leads are introduced into the sample holder, PTFE capsule, through the holed CuBe plug. Plastic fiber (Edmund Optics, Φ 0.25 mm) is introduced together for the *in-situ* pressure calibration using ruby fluorescence methods. The sample is sealed by epoxy resin (STYCAST 2850) immersed with the Cu leads and plastic fiber. If a volume of the epoxy resin is too small compared with the plastic fiber and leads, the epoxy resin cannot keep sample pressure under compression; this means that the sample blows out through the hole. For a similar reason, it is necessary to put epoxy resin on the plug as high as 1.8 mm from the top side of the CuBe plug (this value has been optimized). A small ruby tip (almost the same size of the diameter of plastic fiber) is introduced below the electrode and exposed from the incident 532 nm laser beam travel through the plastic fiber, and the induced fluorescence was also traveled through the fiber to the detector (Ocean optics, USB2000+). The parallel electrodes are fixed on the epoxy resin in the cell assembly. The vertically fixed electrodes ensure a condition that electrode separation and area are constant under compression. It should be noted that the length of the electrode (about 6.0 mm) is an important parameter not to collapse during the compression. If we lengthen the electrodes to expand electrode area (corresponding to enlarge sample capacitance from a principle equation, $C = \varepsilon_0 \varepsilon S/d$), electrode deformation might occur. It may also cause short-circuit between the electrodes. From our experience, the electrode length must be less than 7.0 mm at most, which is the half-length of the initial sample space. The cell assembly allows us to measure dielectric properties of the liquid sample under high pressure.

Supplementary Figure 2 shows an example of dielectric data of non-doped ice VI obtained at 1.73 GPa and 240 K. Supplementary Figure 2a and b show temperature dependence of the dielectric constant and loss of non-doped ice VI obtained at 1.73 GPa, respectively. Similar temperature dependence of dielectric properties of non-doped ice VI was reported by Johari *et al.* at 1.1 GPa[27]. The dielectric response in the lower frequency region (below 10 Hz at 240 K) is derived from the DC electric conductivity of the sample. Temperature dependence of dielectric constants and loss shows that the frequency dispersion derived from the molecular rotation ($10^3$ Hz at 240 K) shifts to lower frequency with decreasing temperature. This means that the dynamics of molecular rotation becomes slowdown with decreasing temperature. As the most important point shown in Supplementary Figure 2, the dielectric response of non-doped ice VI almost disappears at 160 K in the measured frequency range due to the slowdown of the molecular-rotation dynamics. No dielectric response was shown at the hydrogen-ordering temperature of ice VI observed in HCl-doped sample (at ~130 K).

**Supplementary Methods: Details of the 18 candidates**

Supplementary Figure 3 shows definitions of site labels of hydrogen atoms for the 18 structure models. Corresponded site occupancies of the hydrogen atom sites are represented by one or a few variables, denoted by Greek letters (Supplementary Table 1), which are actually fitting parameters for the Rietveld refinements using hydrogen site occupancies.

**Supplementary Note: the criterion of "higher and lower symmetry" space group**

We only took account of 18 candidates of "higher symmetry" space groups for the structure analysis of ice XIX as mentioned in the main text. Considering two structure models, whose space groups are related by group-subgroup relationship, the subgroup structure model generally shows better agreement with the experimental data comparing with the structure model of its parent space group. Hence we set a criterion of "lower symmetry" space groups disregarded in this study based on a quantitative index, Group-Subgroup index. We consider a space group, represented by *G*. *G* is an infinite group due to its translation symmetry. Its translation operations make normal subgroup, $H$, of the space group, $G$. A coset of $G$ by $H$, denoted by $G/H$, makes a quotient group. $G/H$ is composed of finite elements and the number of elements is called order of $G/H$, represented by $|G/H|$. For example, let us take $P1$ as an example space group of a crystal structure, $S$. When we choose a minimum unit cell of the $S$, $T(P1)$ is defined by all translation operations corresponding to the unit cell. Then, $|P1/T(P1)| = 1$ holds. Hereafter, we consider a case that $S$ is the crystal structure of ice VI. In the case of the space group of ice VI, $P4_2/nmc$, $|P4_2/nmc/T(P4_2/nmc)|$ is 16. When the unit cell of ice VI is expanded to $\sqrt{2} \times \sqrt{2} \times 1$, the cell volume becomes two times larger. We represent corresponded translation operations as $T_{\sqrt{2}\times\sqrt{2}\times1}(P4_2/nmc)$. Then an order $|P4_2/nmc/T_{\sqrt{2}\times\sqrt{2}\times1}(P4_2/nmc)|$ is 32. The Group-Subgroup index, $s$, is defined as follows:

$$s = \frac{|P4_2/nmc/T_{\sqrt{2}\times\sqrt{2}\times1}(P4_2/nmc)|}{|H/T(H)|} = \frac{32}{|H/T(H)|},$$

where $H$ is a subgroup of $P4_2/nmc$. For example in a case, $H = P\bar{1}$, Group-Subgroup index, $s$, is 16. Because, $P\bar{1}/T(P\bar{1})$ includes two elements; one is for trivial equivalent symmetry and the other one is about inversion symmetry. In this study, we ignored five space groups, $Pc$, $P2_1$, $P2$ and $P\bar{1}$ and $P1$, and this means that we only considered space groups, whose Group-Subgroup indexes are less than 8 from their sufficient refinement agreements for the neutron diffraction patterns. Space groups with indexes, 16 and 32, are $Pc$, $P2_1$, $P2$ and $P\bar{1}$ and $P1$, respectively. It is noted that subgroups of $P\bar{4}$ are $P2$ and $P1$ both of which belong to pyroelectric groups as well as $Pcc2$.

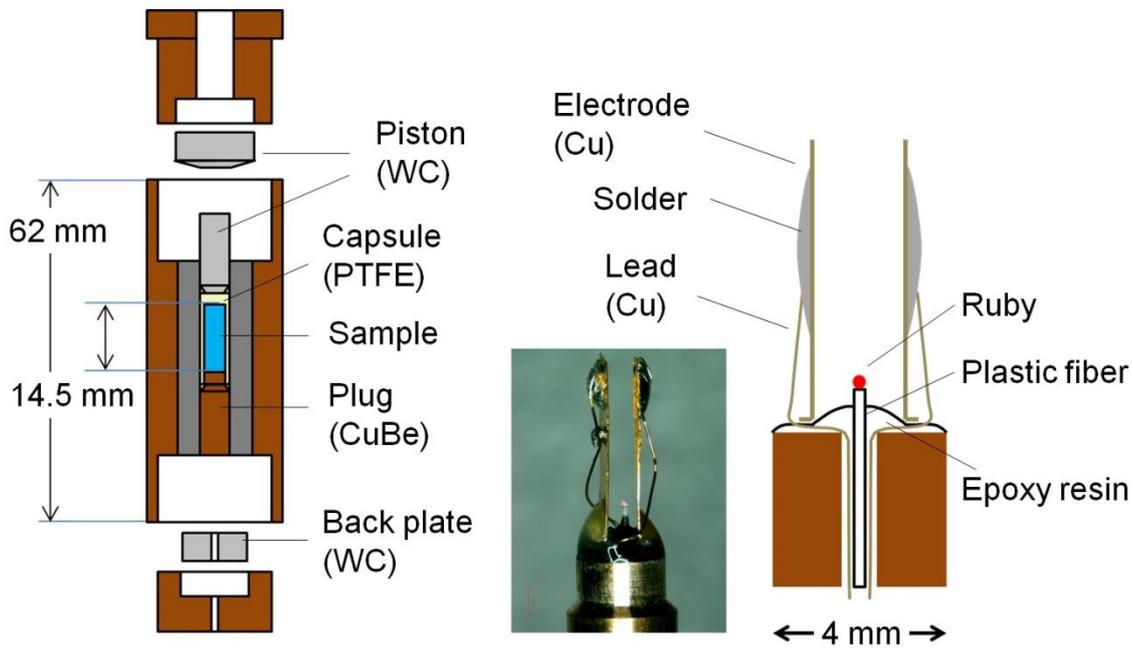

**Supplementary Figure 1| Developed cell assembly of piston cylinder for *in-situ* dielectric measurements under high pressure.** An overall drawing (left) and a picture (center) and schematic drawing around sample space (right) of the developed cell assembly. Separated pistons are made of tungsten carbide abbreviated WC in the figure and cylinder is made of CuBe (outer, colored blown) and NiCrAl (inner, colored gray). The two electrodes compose parallel electrodes.

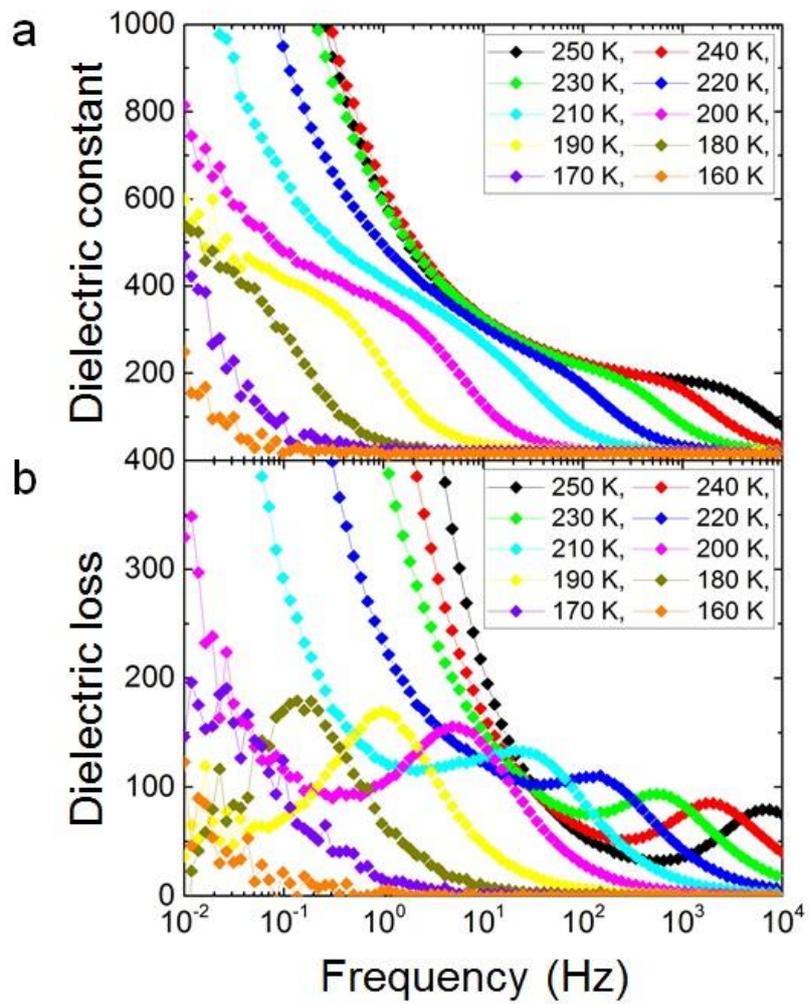

**Supplementary Figure 2| Dielectric properties of non-doped ice VI. a and b**, Temperature dependence of dielectric constant and loss of non-doped ice VI obtained at 1.73 GPa.

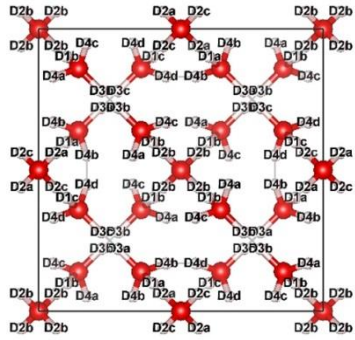
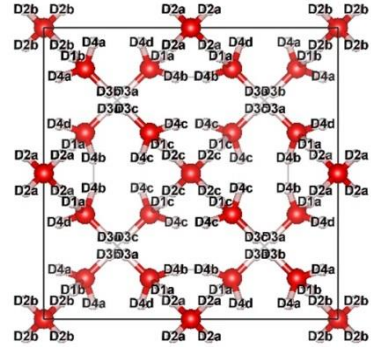
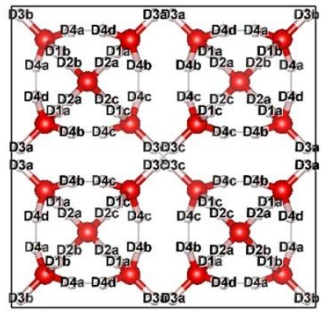
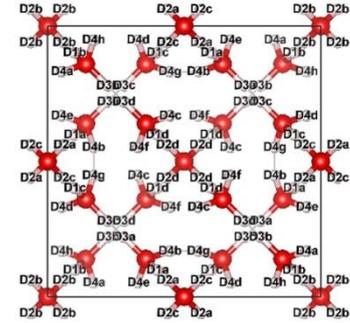
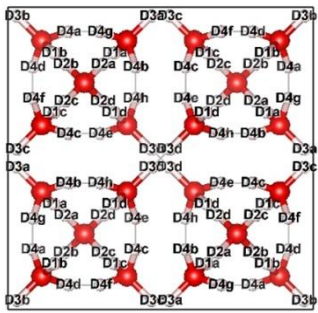
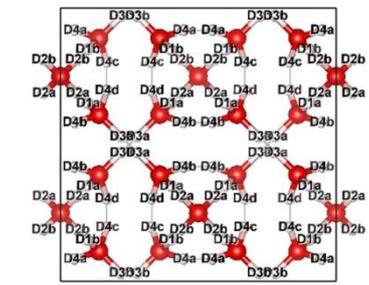

**Supplementary Figure 3| Site labels of hydrogen atoms in the 18 structure models.** Each corresponded space group is denoted left upper of its crystal structure. Black squares mean unit cells and each crystal axis is denoted in the left bottom.

Pba2 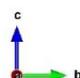 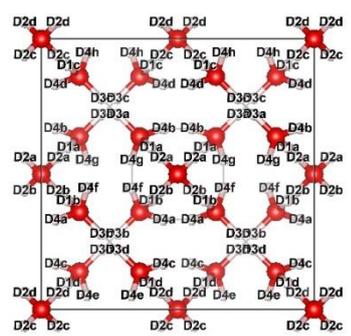 Pca2$_1$ 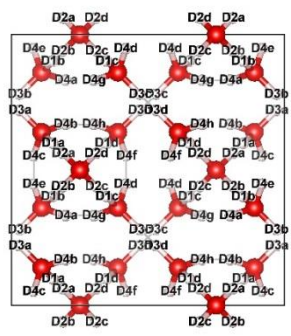

Pcc2 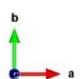 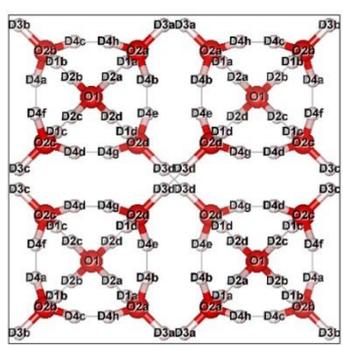 P2$_1$2$_1$2 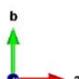 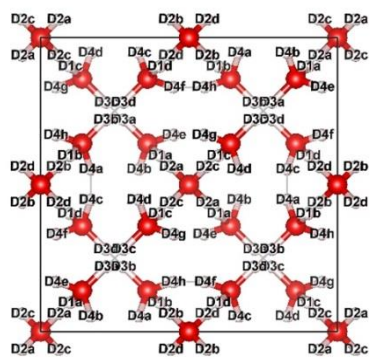

P222$_1$ 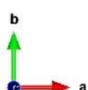 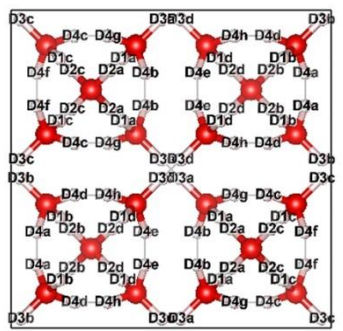 P222 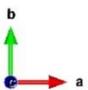 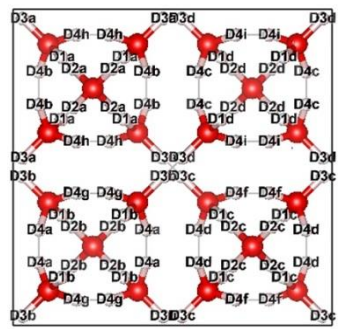

**Supplementary Figure 3| (continuous) Site labels of hydrogen atoms in the 18 structure models.** Each corresponded space group is denoted left upper of its crystal structure. Black squares mean unit cells and each crystal axis is denoted in the left bottom.

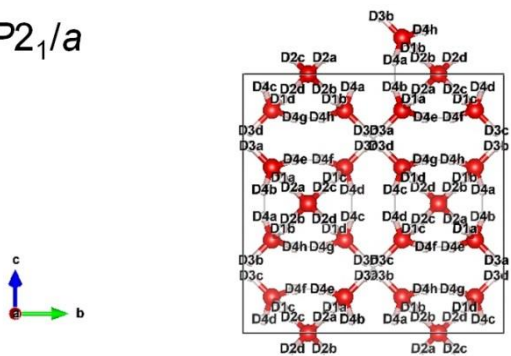
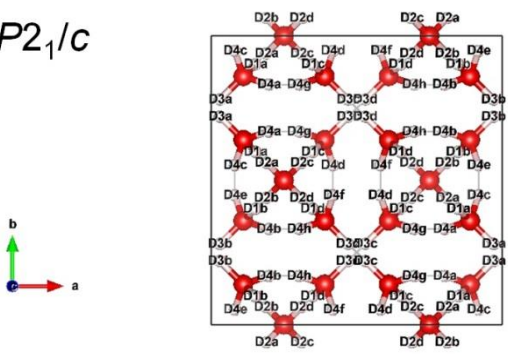
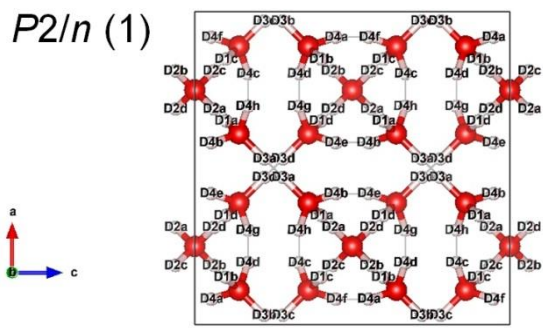
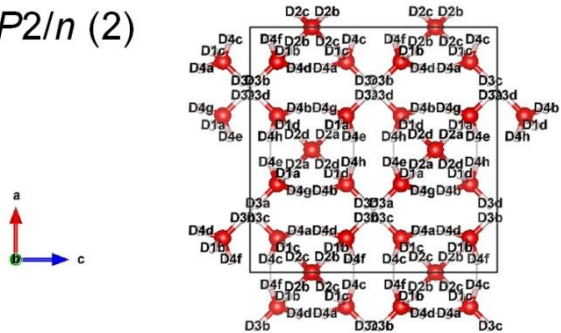
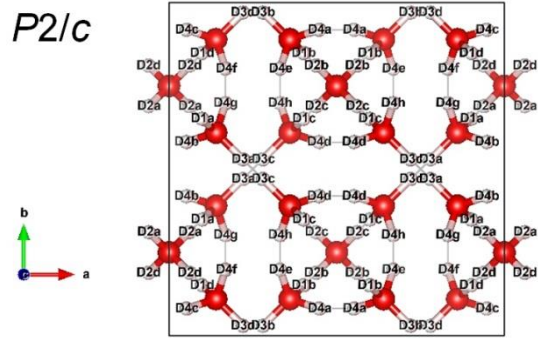
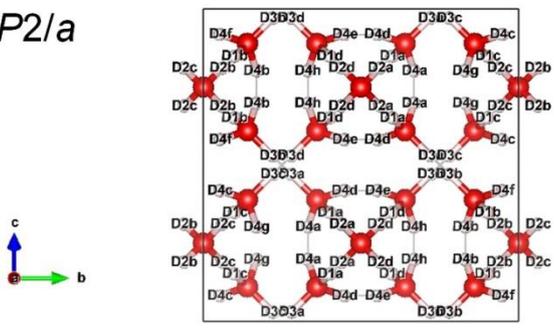

**Supplementary Figure 3| (continuous) Site labels of hydrogen atoms in the 18 structure models.** Each corresponded space group is denoted left upper of its crystal structure. Black squares mean unit cells and each crystal axis is denoted in the left bottom. $P2/n$ has two structure models, denoted by (1) and (2), depending on their geometric elements.

Supplementary Table 1| Representations of hydrogen site occupancies for the 18 structure models

| # | Space group | Site occupancy of each hydrogen atom site |
|---|---|---|
| 1 | $P\bar{4}2_1m$ | D1a : 1 - $\alpha$, D1b : 0.5, D1c : $\alpha$,<br>D2a : $\alpha$, D2b : 0.5, D2c : 1 - $\alpha$,<br>D3a : $\beta$, D3b : 0.5, D3c : 1 - $\beta$,<br>D4a : $\gamma$, D4b : 0.5 + ($\alpha$ - $\beta$)/2,<br>D4c : 1 - $\gamma$, D4d : 0.5 - ($\alpha$ - $\beta$)/2 |
| 2 | $P\bar{4}2m$ | D1a, D1b, D1c : 0.5, D2a, D2b, D2c : 0.5,<br>D3a : 0.5, D3b : $\alpha$, D3c : 1 - $\alpha$,<br>D4a, D4b, D4c, D4d : 0.5 |
| 3 | $P4_2cm$ | D1a : 1 - $\alpha$, D1b : 1 - $\beta$, D1c : -1 + 2$\alpha$ + $\beta$,<br>D2a : $\alpha$, D2b : $\beta$, D2c : 2 - 2$\alpha$ - $\beta$,<br>D3a, D3b, D3c : 0.5,<br>D4a : 0.25 + $\beta$/2, D4b : -0.25 + $\alpha$ + $\beta$/2,<br>D4c : 1.25 - $\alpha$ - $\beta$/2, D4d : 0.75 - $\beta$/2 |
| 4 | $P\bar{4}$ | D1a : 1 - $\alpha$, D1c : $\alpha$, D1b, D1d : 0.5,<br>D2a : $\alpha$, D2c : 1 - $\alpha$, D2b, D2d : 0.5,<br>D3a : $\beta$, D3c : 1 - $\beta$, D3b, D3d : 0.5,<br>D4a : $\gamma$, D4h : 1 - $\gamma$, D4c : $\delta$,<br>D4f : 1 - $\delta$, D4b : $\epsilon$, D4g : 1 - $\epsilon$,<br>D4d : -$\alpha$ + $\beta$ + $\epsilon$, D4e : 1 + $\alpha$ - $\beta$ - $\epsilon$ |
| 5 | $P4_2$ | D1a : 1 - $\alpha$, D1b : 1 - $\beta$, D1c : 1 - $\gamma$, D1d : -1 + $\alpha$ + $\beta$ + $\gamma$,<br>D2a : $\alpha$, D2b : $\beta$, D2c : $\gamma$, D2d : 2 - $\alpha$ - $\beta$ - $\gamma$,<br>D3a, D3b, D3c, D3d : 0.5,<br>D4a : $\delta$, D4d : 0.5 + $\beta$ - $\delta$ D4f : 0.5 - $\beta$ + $\delta$,<br>D4c : $\beta$ + $\gamma$ - $\delta$, D4e : 1 - $\beta$ - $\gamma$ + $\delta$, D4g : 1 - $\delta$,<br>D4b : -0.5 + $\alpha$ + $\delta$, D4h : 1.5 - $\alpha$ - $\delta$ |
| 6 | $Pcca$ | D1a : 1 - $\alpha$, D1b : $\alpha$, D2a : $\alpha$, D2b : 1 - $\alpha$,<br>D3a, D3b : 0.5,<br>D4a, D4b : 0.5, D4c : 1 - $\alpha$, D4d : $\alpha$ |
| 7 | $Pba2$ | D1a : 1 - $\alpha$, D1b : $\alpha$, D1c : 1 - $\beta$, D1d : $\beta$,<br>D2a : $\alpha$, D2b : 1 - $\alpha$, D2c : $\beta$, D2d : 1 - $\beta$,<br>D3a, D3d : $\gamma$, D3b, D3c : 1 - $\gamma$,<br>D4a, D4b, D4c, D4d : 0.5,<br>D4e : 1.5 - $\beta$ - $\gamma$, D4h : -0.5 + $\beta$ - $\gamma$,<br>D4f : 0.5 - $\alpha$ + $\gamma$, D4g : 0.5 + $\alpha$ - $\gamma$ |
| 8 | $Pca2_1$ | D1a : 1 - $\alpha$, D1b : 1 - $\beta$, D1c : 1 - $\gamma$, D1d : -1 + $\alpha$ + $\beta$ + $\gamma$,<br>D2a : $\alpha$, D2b : $\beta$, D2c : $\gamma$, D2d : 2 - $\alpha$ - $\beta$ - $\gamma$,<br>D3a : $\delta$, D3b : 1 - $\delta$, D3c : $\epsilon$, D3d : 1 - $\epsilon$,<br>D4a : $\zeta$, D4g : 1 - $\zeta$, D4e : $\beta$ + $\delta$ - $\zeta$,<br>D4c : 1 - $\beta$ - $\delta$ + $\zeta$, D4b : $\alpha$ + $\beta$ - $\zeta$, D4h : 1 - $\alpha$ - $\beta$ + $\zeta$,<br>D4f : 1 - $\gamma$ + $\epsilon$ - $\zeta$, D4d : $\gamma$ - $\epsilon$ + $\zeta$ |

| 9 | $Pcc2$ | D1a : 1 - $\alpha$, D1b : 1 - $\beta$, D1c : 1 - $\gamma$, D1d : -1 + $\alpha$ + $\beta$ + $\gamma$, D2a : $\alpha$, D2b : $\beta$, D2c : $\gamma$, D2d : 2 - $\alpha$ - $\beta$ - $\gamma$, D3a, D3b, D3c, D3d : 0.5, D4a : $\delta$, D4b : $\alpha$ + $\beta$ - $\delta$, D4c : 0.5 + $\beta$ - $\delta$, D4d : -0.5 + $\gamma$ + $\delta$, D4e : 1 - $\alpha$ - $\beta$ + $\delta$, D4f : 1 - $\delta$, D4g : 1.5 - $\gamma$ - $\delta$, D4h : 0.5 - $\beta$ + $\delta$ |
|---|---|---|
| 10 | $P2_12_12$ | D1a : 1 - $\alpha$, D1b : 1 - $\beta$, D1c : $\alpha$, D1d : $\beta$, D2a : $\alpha$, D2b : $\beta$, D2c : 1 - $\alpha$, D2d : 1 - $\beta$, D3a : $\gamma$, D3b : $\delta$, D3c : 1 - $\gamma$, D3d : 1 - $\delta$, D4a : $\epsilon$, D4c : 1 - $\epsilon$, D4h : 1 + $\beta$ - $\delta$ - $\epsilon$, D4f : -$\beta$ + $\delta$ + $\epsilon$, D4b : $\zeta$, D4d : 1 - $\zeta$, D4e : 1 + $\alpha$ - $\gamma$ - $\zeta$, D4g : -$\alpha$ + $\gamma$ + $\zeta$ |
| 11 | $P222_1$ | D1a : 1 - $\alpha$, D1b : 1 - $\beta$, D1c : $\alpha$, D1d : $\beta$, D2a : $\alpha$, D2b : $\beta$, D2c : 1 - $\alpha$, D2d : 1 - $\beta$, D3a, D3b, D3c, D3d : 0.5, D4a, D4b, D4e, D4f : 0.5, D4c : 1 - $\alpha$, D4g : $\alpha$, D4d : $\beta$, D4h : 1 - $\beta$ |
| 12 | $P222$ | all 0.5 |
| 13 | $P2_1/a$ | D1a : 1 - $\alpha$, D1b : 1 - $\beta$, D1c : 1 - $\gamma$, D1d : -1 + $\alpha$ + $\beta$ + $\gamma$, D2a : $\alpha$, D2b : $\beta$, D2c : $\gamma$, D2d : 2 - $\alpha$ - $\beta$ - $\gamma$, D3a : $\delta$, D3c : 1 - $\delta$, D3b : $\epsilon$, D3d : 1 - $\epsilon$, D4a : $\zeta$, D4b : 1 - $\zeta$, D4e : $\alpha$ - $\delta$ + $\zeta$, D4f : 1 - $\alpha$ + $\delta$ - $\zeta$, D4d : -1 + $\alpha$ + $\gamma$ + $\zeta$, D4c : 2 - $\alpha$ - $\gamma$ - $\zeta$, D4g : -$\beta$ + $\epsilon$ + $\zeta$, D4h : 1 + $\beta$ - $\epsilon$ - $\zeta$ |
| 14 | $P2_1/c$ | D1a : 1 - $\alpha$, D1b : 1 - $\beta$, D1c : 1 - $\gamma$, D1d : -1 + $\alpha$ + $\beta$ + $\gamma$, D2a : $\alpha$, D2b : $\beta$, D2c : $\gamma$, D2d : 2 - $\alpha$ - $\beta$ - $\gamma$, D3a : $\delta$, D3b : 1 - $\delta$, D3c : $\epsilon$, D3d : 1 - $\epsilon$, D4a : $\zeta$, D4g : 1 - $\zeta$, D4d : $\gamma$ - $\epsilon$ + $\zeta$, D4f : 1 - $\gamma$ + $\epsilon$ - $\zeta$, D4h : 1 - $\alpha$ - $\beta$ + $\zeta$, D4b : $\alpha$ + $\beta$ - $\zeta$, D4e : -$\alpha$ + $\delta$ + $\zeta$, D4c : 1 + $\alpha$ - $\delta$ - $\zeta$ |
| 15 | $P2/n$ | D1a : 1 - $\alpha$, D1b : 1 - $\beta$, D1c : 1 - $\gamma$, D1d : -1 + $\alpha$ + $\beta$ + $\gamma$, D2a : $\alpha$, D2b : $\beta$, D2c : $\gamma$, D2d : 2 - $\alpha$ - $\beta$ - $\gamma$, D3a, D3b, D3c, D3d : 0.5, D4a : $\delta$, D4f : 1 - $\delta$, D4c : -0.5 + $\gamma$ + $\delta$, D4h : 1.5 - $\gamma$ - $\delta$, D4b : -1 + $\alpha$ + $\gamma$ + $\delta$, D4e : 2 - $\alpha$ - $\gamma$ - $\delta$, D4g : 0.5 - $\beta$ + $\delta$, D4d : 0.5 + $\beta$ - $\delta$ |
| 16 | $P2/n$ | D1a, D2d : 1 - $\alpha$, D1b, D2c : 1 - $\beta$, D2a, D1d : $\alpha$, D2b, D1c : $\beta$, D3a, D3c : $\gamma$, D3b, D3d : 1 - $\gamma$, D4a : $\delta$, D4d : 1 - $\delta$, D4f : -1 + $\beta$ + $\gamma$ + $\delta$, D4c : 2 - $\beta$ - $\gamma$ - $\delta$, D4b : $\epsilon$, D4g : 1 - $\epsilon$, D4e : $\alpha$ - $\gamma$ + $\epsilon$, D4h : 1 - $\alpha$ + $\gamma$ - $\epsilon$ |
| 17 | $P2/c$ | D1a, D2d : 1 - $\alpha$, D1b, D2c : 1 - $\beta$,<br>D2a, D1d : $\alpha$, D2b, D1c : $\beta$,<br>D3a, D3b : $\gamma$, D3c, D3d : 1 - $\gamma$,<br>D4a, D4b, D4c, D4d : 0.5,<br>D4e : 0.5 + $\beta$ - $\gamma$, D4h : 0.5 - $\beta$ + $\gamma$,<br>D4f : 0.5 - $\alpha$ + $\gamma$, D4g : 0.5 + $\alpha$ - $\gamma$ |
| 18 | $P2/a$ | D1a, D2d : 1 - $\alpha$, D1b, D2c : 1 - $\beta$,<br>D2a, D1d : $\alpha$, D2b, D1c : $\beta$,<br>D3a, D3c : $\gamma$, D3b, D3d : 1 - $\gamma$,<br>D4a, D4b, D4g, D4h : 0.5,<br>D4c : 1.5 - $\beta$ - $\gamma$, D4f : -0.5 + $\beta$ + $\gamma$,<br>D4d : 0.5 + $\alpha$ - $\gamma$, D4e : 0.5 - $\alpha$ + $\gamma$ |